\documentclass[preprint,showpacs,titlepage,aps,prd,
amsmath,byrevtex,nofootinbib]{revtex4}

\usepackage{graphicx}

\begin{document}

\preprint{FERMILAB-PUB-04-140-T}
\preprint{hep-ph/0408070}
\vspace*{-2cm}

\title{\Large 
Untangling\\ CP 
Violation and the Mass 
Hierarchy\\ in Long 
Baseline Experiments}

\author{{\large Olga Mena and Stephen Parke}}

\vspace*{1.0cm}
\affiliation{{\sl Theoretical 
Physics Department,
Fermi National 
Accelerator Laboratory 
\\
P.O.Box 500, Batavia, 
IL 60510, USA}\\
omena@fnal.gov 
parke@fnal.gov
}
\date{July 31, 2004\\[1cm]}

\vglue 1.4cm

%%%%%%%%%%%%%%%%%%%%%%%%%%%%%%%%%%%%%%%%%%%%%%%%%%%%%%%%%%%%%%%%%
%    Abstract
%%%%%%%%%%%%%%%%%%%%%%%%%%%%%%%%%%%%%%%%%%%%%%%%%%%%%%%%%%%%%%%%%
%\hfuzz=25pt
\begin{abstract}
In the overlap region, for the normal and inverted hierarchies, of the
neutrino-antineutrino bi-probability space for $\nu_\mu \to \nu_e$
appearance, we derive a simple identity between the solutions in the
($\sin^2 2\theta_{13}$, $\sin \delta$) plane for the different
hierarchies. The parameter $\sin^2 2\theta_{13}$ sets the scale of the
$\nu_\mu \to \nu_e$ appearance probabilities at the atmospheric $\delta
m^2_{atm} \approx 2.4 \times 10^{-3}$ eV$^2$ whereas $\sin \delta $
controls the amount of CP violation in the lepton sector. The identity
between the solutions is that the difference in the values of $\sin \delta
$ for the two hierarchies equals twice the value of $\sqrt{\sin^2
2\theta_{13}}$ divided by the {\it critical} value of $\sqrt{\sin^2
2\theta_{13}}$. We apply this identity to the two proposed long baseline
experiments, T2K and NO$\nu$A, and we show how it can be used to provide a
simple understanding of when and why fake solutions are excluded when 
two or more experiments are combined.  This identity demonstrates the true
complimentarity of T2K and NO$\nu$A.
\end{abstract}

\maketitle

%%%%%%%%%%%%%%%%%%%%%%%%

With the possibility of the first measurement of $\theta_{13}$
being made by a 1 to 2 km baseline reactor experiment~\cite{reactor_expt}, the
long baseline $\nu_e$ appearance experiments, T2K~\cite{t2k} and NO$\nu$A~\cite{nova}, 
need to adjust their focus to emphasize other
physics topics. 
The most important of these questions is the form of the mass hierarchy, normal
($\delta m^2_{31} > 0$)
versus inverted ($\delta m^2_{31}<0$), 
and whether or not leptonic CP violation occurs, ($\sin \delta \neq 0$).
Matter effects~\cite{matter} entangle these questions~\cite{matterosc}.
Suppose $P(\nu_\mu \to \nu_e) <
P(\bar{\nu}_\mu \to \bar{\nu}_e)$, then in vacuum this implies CP violation, however 
in matter this implies CP violation only for the normal hierarchy
but not necessarily for the inverted hierarchy.
The purpose of this paper is to demonstrate that there is a simple way to understand
this entanglement and to use this understanding to untangle the mass hierarchy
question from whether or not leptonic CP violation occurs.

The outline of this paper is as follows:   Along the diagonal of the $\nu_\mu \to \nu_e$
bi-probability diagram, see Figs. 1 and 2, we solve for $\theta_{13} ~{\rm and}~ \delta$
exactly, i.e. we have imposed the constraint
$P(\nu_\mu \to \nu_e) =P(\bar{\nu}_\mu \to \bar{\nu}_e)$.
There are four such 
solutions\footnote{We assume $\theta_{23}=\pi/4$~\cite{sk,lisi} initially and discuss generalizations later.},
two for the normal hierarchy~\cite{deg1} and two for the inverted hierarchy~\cite{deg2,deg3}.
With these solutions we derive an identity connecting the difference in the
mean values of $\sin \delta$ (the CP violating parameter) for the two hierarchies
to the mean values of $\theta_{13}$ for these solutions.
Although this identity is derived along the diagonal, in an Appendix we present the corrections
to this identity off the diagonal using the approximate solutions derived in Ref.\cite{mnp2}.
We then apply this identity to the proposed long baseline experiments T2K and NO$\nu$A.
We show that the fake solutions for these two experiments 
occur in different parts of parameter
space and therefore they can be excluded with sufficient statistics~\cite{prevst}.
The identity relating the two mean values of $\sin \delta$, one for the normal hierarchy
and one for inverted hierarchy is the new result of this paper and 
it provides a simple physics understanding of when various fake solutions 
are excluded when experiments are combined.

The $\nu_\mu \to \nu_e$ appearance probabilities in long baseline neutrino
oscillation experiments, assuming the normal mass hierarchy, can be written as~\cite{deg1}

\begin{eqnarray}
P(\nu_\mu\to \nu_e)& = & X_+ \theta^2 +Y_+ \theta \cos(\Delta_{13}+\delta) + P_\odot \nonumber \\
\overline{P}(\bar{\nu}_\mu \to \bar{\nu}_e) & = & X_-  \theta^2 -Y_-  \theta  \cos(\Delta_{13}-\delta)
+P_\odot  .
\label{e_appear}
\end{eqnarray}
In the last expressions,  $\theta=\sin \theta_{13}$ and the coefficients $X_{\pm}$ and $Y_{\pm}$ are determined by
\begin{eqnarray}
X_{\pm} &=& 4 s^2_{23}
\left\{ \frac{\Delta_{13}\sin({aL \mp \Delta_{13}})}{(aL \mp \Delta_{13})} \right\}^2 , 
\nonumber  \\
Y_{\pm} &=& \pm 2\sqrt{X_\pm P_\odot} =\pm 8 c_{12}s_{12}c_{23}s_{23}
\left\{ \frac{\Delta_{13}\sin({aL \mp \Delta_{13}})}{(aL \mp \Delta_{13})} \right\}
\left\{ \frac{\Delta_{12}\sin{({aL})}}{aL}\right\}
\label{Y}\\
P_{\odot} & = & c^2_{23} \sin^2{2\theta_{12}} \left\{ \frac{\Delta_{12}\sin{({aL})}}{aL}\right\}^2
\nonumber
\end{eqnarray}
where $\Delta_{ij}  \equiv |\Delta m^2_{ij}| L/4E$ and $a = G_F N_e/\sqrt{2}$ denotes the index of refraction 
in matter with $G_F$ being the Fermi constant and $N_e$ a constant 
electron number density in the earth. 
Obviously from the above definitions, 
$X_\pm$ and $Y_\pm$ satisfy the identity
\begin{equation}
\frac{Y_+}{\sqrt{X_+}}
= - ~\frac{Y_-}{\sqrt{X_-}}
\end{equation}
which is used extensively throughout this paper.

To solve equations Eqn.[\ref{e_appear}] 
exactly with the constraint $P=\overline{P}$, 
i.e. along the diagonal of the bi-probability diagram, 
we use the ansatz%\footnote{Away from the diagonal multiply the $P$ equation by 
%$\overline{P}$ and $\overline{P}$ equation by $P$ and proceed as before for 
%general solution.}
\begin{eqnarray}
\theta & = & \theta_c (\sin \delta - \beta \cos \delta)
\end{eqnarray}
where
\begin{eqnarray}
\theta_c = \frac{Y_+}{\sqrt{X_+}} { \sin \Delta_{13} \over (\sqrt{X_+} - \sqrt{X_-})} & \quad {\rm and}
\quad & \beta = \left( {\sqrt{X_+} - \sqrt{X_-}   \over \sqrt{X_+} +\sqrt{X_-}  } \right) 
\frac{\cos \Delta_{13}}{\sin \Delta_{13}}.
\end{eqnarray}
Then
\begin{eqnarray}
P=\overline{P} & = & \sqrt{X_+} \sqrt{X_-} ~ \theta_c^2 ~(\sin^2 \delta - \beta^2 \cos^2 \delta)+P_\odot.
\end{eqnarray}
P has a maximum when $\sin \delta =1$, $\theta=\theta_c$ and 
$P_c =\sqrt{X_+} \sqrt{X_-} ~ \theta_c^2+P_\odot$.  We call these values 
the critical values of P and $\theta$.
There are no solutions along the diagonal for values of P larger than $P_c$.

Using this critical value of P to normalize the probabilities, we can solve for $\delta$. Thus the 
exact solutions, labeled 1 and 2, for the normal hierarchy, are
\begin{eqnarray}
\theta_1=\theta_c~(s_p - \beta c_p), \quad &\sin \delta_1 = s_p    \quad {\rm and} & \cos \delta_1= c_p
\nonumber \\
\theta_2=\theta_c~(s_p + \beta c_p), \quad &\sin \delta_2 = s_p    \quad {\rm and} & \cos \delta_2=-c_p
\end{eqnarray}
where
\begin{eqnarray}
s_p \equiv +\sqrt{ (P-P_\odot)/(P_c-P_\odot) +\beta^2 \over 1+\beta^2}   \quad &{\rm and}& \quad 
c_p \equiv +\sqrt{1-(P-P_\odot)/(P_c-P_\odot) \over 1+ \beta^2}.
\end{eqnarray}
Along the diagonal the  two solutions for the CP violating parameter, $\sin \delta$,
are identical, $\sin \delta_1 =\sin \delta_2$.

For the inverted hierarchy, the $\nu_\mu \to \nu_e$ appearance probabilities are  
\begin{eqnarray}
P(\nu_\mu\to \nu_e) & = & X_-  \theta^2 +Y_-  \theta \cos(\Delta_{13}-\delta) +P_\odot
\nonumber \\
\overline{P}(\bar{\nu}_\mu \to \bar{\nu}_e) & = & X_+  \theta^2 -Y_+  \theta \cos(\Delta_{13}+\delta)
+P_\odot.
\end{eqnarray}
These equations are identical to the equations for the normal hierarchy when we use the constraint
$P=\overline{P}$ and replace $\delta$ with $ \delta + \pi$, then, the solutions are
\begin{eqnarray}
\theta_3=\theta_c~(s_p - \beta c_p), \quad &\sin \delta_3 = -s_p    \quad {\rm and} & \cos \delta_3= -c_p
\nonumber \\
\theta_4=\theta_c~(s_p + \beta c_p), \quad &\sin \delta_4 = -s_p    \quad {\rm and} & \cos \delta_4=c_p.
\end{eqnarray}
Note that $\theta_3 = \theta_1$ with $\delta_3=\pi+\delta_1$ and $\theta_4 = \theta_2$ with 
$\delta_4=\pi+\delta_2$.

With these solutions in hand it is simple to derive the principal result of this paper,
{
\mathversion{bold}
\begin{eqnarray}
 \langle \sin \delta \rangle_+
- \langle \sin \delta \rangle_- 
%& = & { 2 \langle \theta \rangle \over \theta_c } 
& = &  { 2 \langle \theta \rangle / \theta_c }  
% \\ | \sin \delta_{2i} -\sin \delta_{2i-1} | & = & 0 \quad ~{\rm for} ~i=1,2
\label{mp_master}
\end{eqnarray}
}
where  $\langle \sin \delta \rangle_{+(-)} = (\sin \delta_{1(3)}+\sin \delta_{2(4)})/2$,
the mean values of $\sin \delta$ for each hierarchy, and 
 $\langle \theta \rangle = (\theta_1 +\theta_2+\theta_3+\theta_4)/4$, the mean value 
 of $\theta$ for both hierarchies. 
 For $P=\overline{P}$ there are many ways to write this expression,  however we write it
in this way because with these variables it is accurate even if $P\neq \overline{P}$.
In vacuum, $\theta_c \to \infty$ so that the values of $\sin \delta$ for the two hierarchies
are identical.

The physical meaning of this result is clear, i.e the difference in the mean values
of $\sin \delta$ (the CP violating parameter) between the mass hierarchies
equals twice the mean value of $\theta$ divided by the critical value of $\theta$.
Away from $P=\overline{P}$ it is well known that the difference between the solutions
for $\sin \delta$ and $\theta$ within the same hierarchy are small\cite{prevst}. 
This implies that the relationship given by Eqn.[\ref{mp_master}]  is 
still useful and informative even when $P \neq \overline{P}$.
In fact we have used the approximations of Ref.\cite{mnp2} to derive the corrections to
this master equation and find that the corrections are of ${\cal O}(\beta^2)$. Also
the difference between the solutions of $\sin \delta$ within a hierarchy are of
${\cal O}(\beta)$, see the Appendix.
For the currently proposed experiments $\beta$ is less than or of order 0.1 so the corrections
to Eqn.[\ref{mp_master}] are no larger than a few percent.
In a follow up paper, we will explore in more detail the accuracy of this relationship throughout
the whole overlap region.

The proposed long baseline, off-axis experiments are T2K and No$\nu$A. 
T2K utilizes a steerable neutrino beam from JHF and SuperKamiokande and/or HyperKamiokande as the far detector.
The mean energy of the neutrino beam will be tuned to be at vacuum oscillation maximum,
$\Delta_{13}= \frac{\pi}{2}$, which implies a $\langle E_\nu \rangle =0.6$ GeV at the baseline of  295 km using $|\delta m^2_{31} |= 2.4 \times 10^{-3}$eV$^2$~\cite{sk}. This is the 3$^o$ off-axis beam. 
For this configuration the matter effects are small but not neglible~\cite{matter_t2k} as can be seen
from the separation of the allowed regions in the bi-probability diagram, Fig. 1, for this experiment.
Applying our identity, Eqn.[\ref{mp_master}], to T2K, we find:  
{
\mathversion{bold}
\begin{eqnarray}
 \langle \sin \delta \rangle_+
- \langle \sin \delta \rangle_- 
%& = & { 2 \langle \theta \rangle \over \theta_c } 
& = &  0.47 \sqrt{\sin^2 2\theta_{13} \over 0.05}  \quad \quad \quad {\rm for ~T2K}
\end{eqnarray}
}
i.e. the difference between the true and fake solutions for the
CP violating parameter $\sin \delta$ is  0.47 ($\approx \sqrt{2}/3$) at $\sin^2 2\theta_{13} = 0.05$.

NO$\nu$A proposes to use the Fermilab NuMI beam with a baseline of 
810 km  with a 50 kton low Z detector which is 10km off-axis resulting in
a mean neutrino energy of 2.3 GeV.  The NO$\nu$A beam energy is about 30\%
above the vacuum oscillation maximum energy for this baseline.
Matter effects are quite significant for NO$\nu$A as can be seen from the 
bi-probability diagram, Fig 2. Applying our identity to NO$\nu$A we find:
{
\mathversion{bold}
\begin{eqnarray}
 \langle \sin \delta \rangle_+
- \langle \sin \delta \rangle_- 
%& = & { 2 \langle \theta \rangle \over \theta_c } 
& = &  1.41 \sqrt{\sin^2 2\theta_{13} \over 0.05}  \quad \quad \quad {\rm for ~NO\nu A}.
\end{eqnarray}
}
The difference between the true and fake solutions for the CP violating parameter
$\sin \delta$ is  1.41 ($\approx \sqrt{2}$) at $\sin^2 2\theta_{13} = 0.05$.
The factor of 3 increase in the difference of the $\sin \delta$'s compared to T2K 
is due to the coefficient
in front of the square root which is proportional to (aL).  
The NO$\nu$A detector is 2.75 times further
away from the source than the T2K detector 
and the average density for the NOVA baseline is slightly higher than 
for the T2K baseline.

Combining the results from T2K and NO$\nu$A we note that for the correct hierarchy
and hence the true value of $\sin \delta$ the results should coincide within uncertainties
{
\mathversion{bold}
\begin{eqnarray}
| ~\langle \sin \delta \rangle_{true}^{T2K}
- \langle \sin \delta \rangle_{true}^{NO\nu A}~|
%& = & { 2 \langle \theta \rangle \over \theta_c } 
& \approx &  0.
\end{eqnarray}
}
Whereas for the wrong hierarchy, the fake solutions of 
 $\sin \delta$ are separated by
{
\mathversion{bold}
\begin{eqnarray}
| ~\langle \sin \delta \rangle_{fake}^{T2K}
- \langle \sin \delta \rangle_{fake}^{NO\nu A}~|
%& = & { 2 \langle \theta \rangle \over \theta_c } 
& = &  0.94 \sqrt{\sin^2 2\theta_{13} \over 0.05} .
\end{eqnarray}
}
This implies that if $\sin \delta$ can be measured with sufficient accuracy in both
experiments not only could the hierarchy be determined but also the  true value of
the CP violating parameter $\sin \delta$ including  in the overlap region.
Even for $\sin^2 2\theta_{13} = 0.01$, the separation of the fake solutions
of $\sin \delta$ between experiments is 0.40.  

In Figs. 3 and 4 we have constructed the 
$\chi^2$ 
contours for both T2K and NO$\nu$A
assuming that the true solution is the normal hierarchy and that the values of 
($\sin^2 2 \theta_{13}$, $\delta$) are (0.05, $\pi/4$) respectively.
This point is near the middle of the overlap region in the bi-probability diagram for both T2K and NO$\nu$A
and it is one of the harder points to untangle the mass hierarchy and determine CP violation.
Since T2K is operated at vacuum oscillation maximum there are only two
allowed regions in the $(\sin^2 2\theta_{13},~\sin \delta)$ plane since this 
experiment is 
insensitive to the CP conserving
quantity $\cos \delta$.  NO$\nu$A on the other hand is operated above oscillation
maximum so this experiment is sensitive to the sign\footnote{Given $\sin \delta$ one knows
the magnitude of $\cos \delta$.}  of $\cos \delta$. 
Therefore there are four solutions in  $(\sin^2 2\theta_{13},~\sin 
\delta)$ plane.
The approximate exposure that makes the ellipses in Figs. 3 and 4 the 68, 90 and 99\% C.L. contours is 5 years of both neutrino and anti-neutrino running with T2K operating at 0.75MW using HyperKamiokande as the detector and No$\nu$A operating at 2 MW with a 50kton low Z detector.\footnote{We choose this combination so that the statistical uncertainty in $\sin \delta$ is approximately the same for both experiments.} 
Clearly, when the results of these two experiments are combined
only the region near the true solution (normal hierarchy, $\sin^2 2\theta_{13} \approx 0.05$ and
$\sin \delta \approx 0.7$ and $\cos \delta > 0$), survives at more than 99\% C.L.

If we allow $\theta_{23}$ to vary from $\pi/4$ then the best variables to use are
$\sqrt{2} \cos \theta_{23} \sin \delta$ and  $2 \sin^2 \theta_{23} \sin^2 2\theta_{13}$.
Using these variables we obtain the following identities:
{
\mathversion{bold}
\begin{eqnarray}
 \sqrt{2} \cos \theta_{23} 
 \langle 
\sin \delta \rangle_+
- \sqrt{2} \cos \theta_{23} \langle \sin \delta \rangle_-   
%& = & { 2 \langle \theta \rangle \over \theta_c } 
& = &  0.47 \sqrt{ 2 \sin^2 \theta_{23} \sin^2 2\theta_{13} \over 0.05}   \quad {\rm for ~T2K}\nonumber \\
& & \\ %[0.5cm]
\sqrt{2} \cos \theta_{23} 
\langle \sin \delta \rangle_+
- \sqrt{2} \cos \theta_{23} \langle \sin \delta \rangle_- 
%& = & { 2 \langle \theta \rangle \over \theta_c } 
& = &  1.41 \sqrt{2 \sin^2 \theta_{23} 
\sin^2 2\theta_{13} \over 0.05}  \quad {\rm for ~NO\nu A}. \nonumber
\end{eqnarray}
}
With these variables the figures equivalent
to Figs. 3 and 4 but with $\sin^2 \theta_{23}$ varying between
0.35 and 0.65 (the allowed region from SuperKamiokande 
atmospheric neutrino results~\cite{sk}) are almost identical
except near the upper and lower boundary since the range of  $\sqrt{2} \cos \theta_{23} \sin \delta$
for fixed $\sin^2 \theta_{23}$ is $\pm \sqrt{2} \cos \theta_{23}$, not $\pm1$ as it is for 
$\theta_{23} =\pi/4$.

In summary we have derived a simple identity relating the solutions between the two hierarchies
which allows one to compare the results from two or more long baseline experiments in a very straight
forward manner.  This identity was applied to the proposed T2K and NO$\nu$A experiments
and it demonstrates the true complimentary of these experiments in a simple, transparent fashion.

\section{Acknowledgements}
 The authors wish to thank Hisakazu Minakata, Hiroshi Nunokawa, Takaaki Kajita and Mark Meisser
 for discussions. 
 Fermilab is operated under DOE contract DE-AC02-76CH03000.
  
\newpage
\section{Appendix}
For $P\neq \overline{P}$ we use the solutions, notation and approximations of \cite{mnp2}:
(1 and 2 are labels for the solutions for the
 normal hierarchy and 3 and 4 for the inverted hierarchy.)\\
If we define 
\begin{eqnarray}
\langle \sin \delta \rangle_+ & \equiv & (\sin \delta_1 +\sin \delta_2)/2 \\
\langle \sin \delta \rangle_- & \equiv & (\sin \delta_3 +\sin \delta_4)/2 \\
\langle \theta \rangle & \equiv & (\theta_1+\theta_2+\theta_3+\theta_4)/4\\[0.25cm]
\Omega \equiv 1+\beta^2 &=& 1+ { (\sqrt{X_+} -\sqrt{X_-})^2 \cos^2 \Delta
\over (\sqrt{X_+} +\sqrt{X_-})^2 \sin^2 \Delta} \approx 1,
\end{eqnarray}
then from Eqn. (34)-(37) of \cite{mnp2} we find
\begin{eqnarray}
\langle \sin \delta \rangle_+ - \langle \sin \delta \rangle_-
 & = & 2 \left\{ \sqrt{P} +\sqrt{\overline{P}} \over \sqrt{X_+} +\sqrt{X_-} \right\}
 \left\{ \sqrt{X_+}(\sqrt{X_+}-\sqrt{X_-}) \over Y_+ \sin \Delta \right\} 
 \Omega^{-1},\\
%\end{eqnarray} 
%\begin{eqnarray}
\langle \theta \rangle & = & \left\{ \sqrt{P} +\sqrt{\overline{P}} \over \sqrt{X_+} +\sqrt{X_-} \right\}
\Omega^{-1}\\
%\end{eqnarray}
{\rm and} \quad \quad
%\begin{eqnarray}
\theta_{crit} & = &\left\{Y_+ \sin \Delta \over \sqrt{X_+}(\sqrt{X_+}-\sqrt{X_-})  \right\} \Omega^{-\frac{1}{2}}.
\end{eqnarray}
These solutions therefore satisfy
\begin{eqnarray}
\langle \sin \delta \rangle_+ - \langle \sin \delta \rangle_-
 & = & 2 ~\Omega^{-\frac{1}{2}}~\langle \theta \rangle / \theta_{crit} 
% \approx 2 ~\langle \theta \rangle / \theta_{crit}.
\end{eqnarray}
throughout the overlap region. This identity is identical  to Eqn.[\ref{mp_master}] up to small corrections.

%For consistency, note that at the critical point 
%$\langle \sin \delta \rangle_+ = -\langle \sin \delta \rangle_- =\Omega^{-\frac{1}{2}}$.\\
%and that the $P_{crit}$ given by eqn(49) of \cite{mnp2}  can be rewritten as
%\begin{eqnarray}
%P_{crit} & = & \Omega \left( Y_+ \over 2 \sqrt{X_+} \right)^2 \left( \sqrt{X_+} +\sqrt{X_-} \over
%\sqrt{X_+} - \sqrt{X_-} \right)^2 \sin^2\Delta.
%\end{eqnarray}
%Substitution this into the expression for  $\langle \theta \rangle$ 
%gives $\theta_{crit}$ above.

This identity is only useful and informative if both $|\theta_i-\theta_j|$ and $|\sin \delta_i - \sin \delta_j | $
for (i,j)= (1,2) or (3,4) are small i.e. in the same hierarchy.  From the solutions in Ref.\cite{mnp2}, one can easily derive that
\begin{eqnarray}
|\theta_i-\theta_j| &\leq& \beta \theta_{crit}
=\begin{cases}
  ~\approx 0  & \text{T2K},\\
 ~\leq 0.02 & \text{NO$\nu$A.}
 \end{cases}
\end{eqnarray}
For NO$\nu$A this restricts the usefulness of our identity to $\sin^2 2\theta_{13} > 10^{-3}$.

The difference between the two values of $\sin \delta$ in the SAME hierarchy from
Eqn.(34) and (35) of Ref.\cite{mnp2} is bounded by 
\begin{eqnarray}
|\sin \delta_i - \sin \delta_j | & \leq & \beta={ (\sqrt{X_+} -\sqrt{X_-}) \cos \Delta
\over (\sqrt{X_+} +\sqrt{X_-}) \sin \Delta}  \quad {\rm for~(i,j) = (1,2) ~or ~(3,4)}\\
 & \approx & (aL)(\Delta^{-1} - \cot \Delta)\cot \Delta 
=\begin{cases}
  ~\approx 0  & \text{T2K},\\
 ~\leq 0.1 & \text{NO$\nu$A.}
 \end{cases}
\end{eqnarray}
for (i,j) = (1,2) or (3,4). 

In conclusion, the identity presented in this paper is accurate, useful and informative for
all values of the parameters that can be probed by the proposed experiments T2K and NO$\nu$A.
For very small values of $\theta_{13}$, beyond the reach of these experiments, there can be significant
corrections but here the separation of the $\sin \delta$'s between the hierarchies is small.

\clearpage
\begin{figure}[h]
\begin{center}
\includegraphics[width=4in]{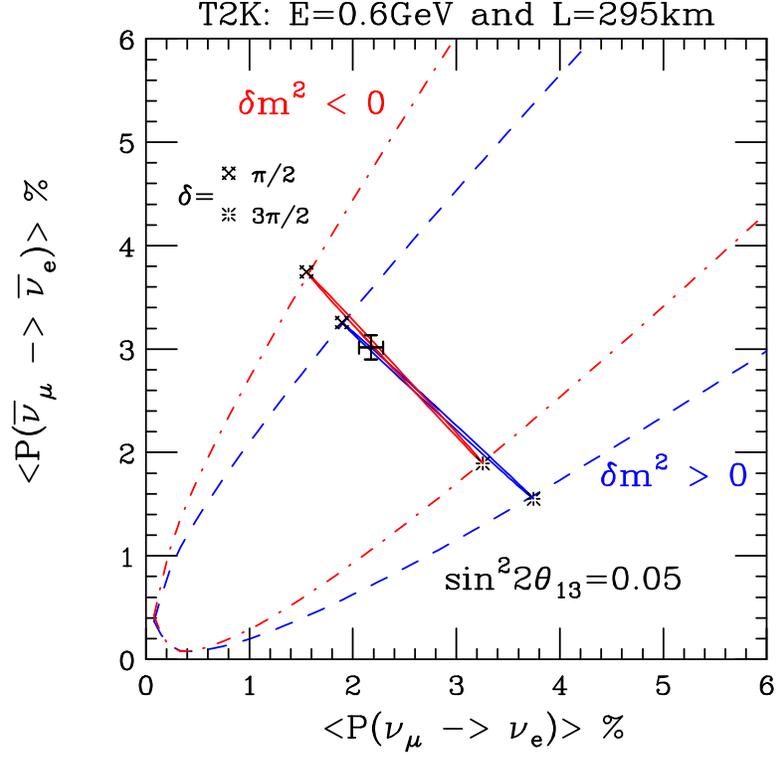}
%\vspace*{-2.5in}
\caption[]{The bi-probability diagram for T2K showing the allowed regions for both
the normal (dashed) and inverted (dotdashed) hierarchies as well as the ellipses for $\sin^2 2\theta_{13} =0.05$.
The large ``{\large $+$}'' marks the neutrino and anti-neutrino probabilities with the CP phase, $\delta = \pi/4$, assuming the normal hierarchy. The critical value for this experiment is way off this figure.}
\end{center}
\end{figure}

\clearpage
\begin{figure}[h]
\begin{center}
\includegraphics[width=4in]{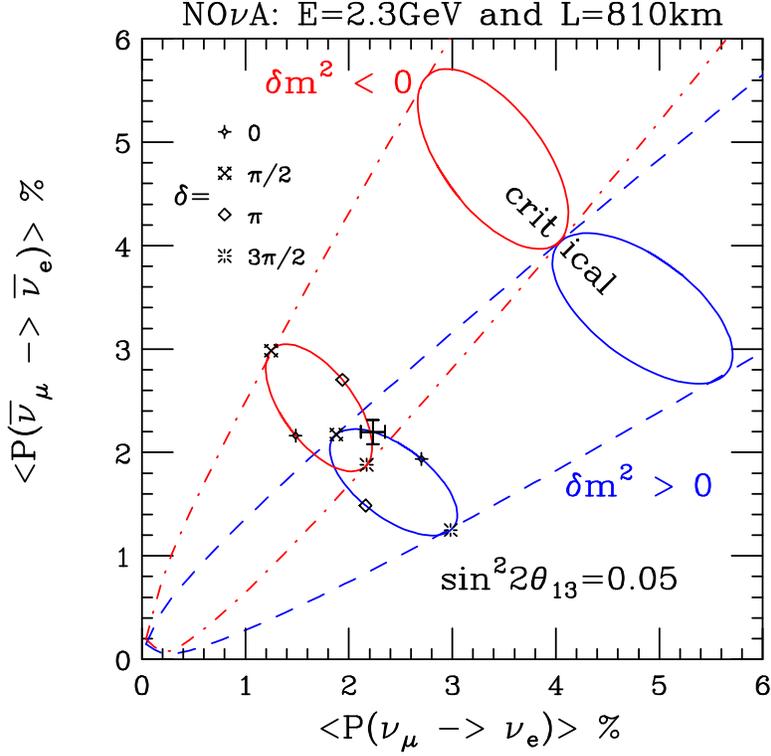}
%\vspace*{-2.5in}
\caption[]{The bi-probability diagram for NO$\nu$A showing the allowed regions for both
the normal (dashed) and inverted (dotdashed) hierarchies as well as the ellipses for $\sin^2 2\theta_{13} =0.05$.
The large ``{\large $+$}'' marks the neutrino and anti-neutrino probabilities with the CP phase, $\delta = \pi/4$, assuming the normal hierarchy. The ellipses and point along the diagonal
labeled critical correspond to the largest values for which there is overlap between the normal
and inverted hierarchies. }
\end{center}
\end{figure}
\clearpage

\begin{figure}[h]
\begin{center}
\includegraphics[width=4in]{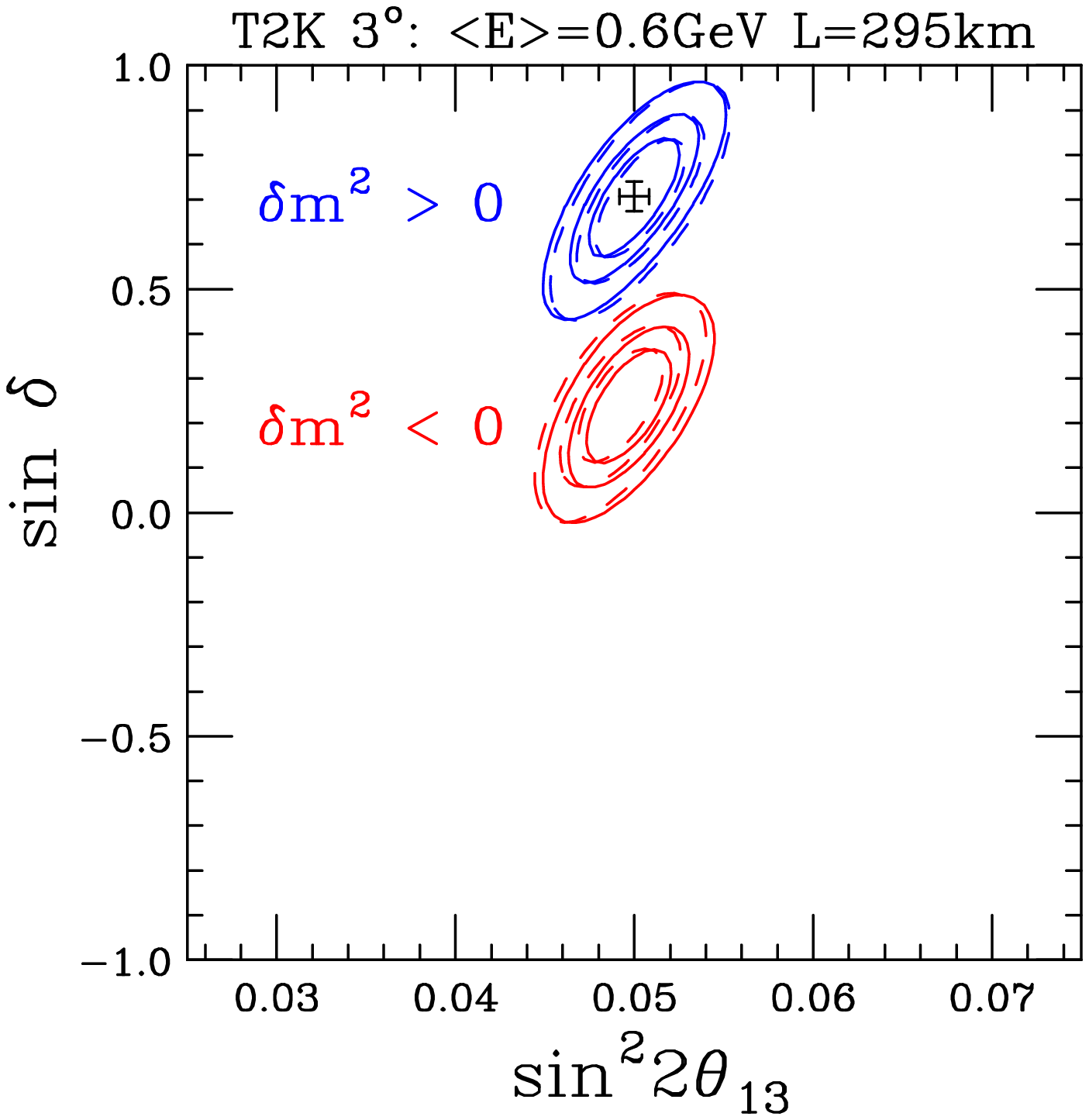}
%\vspace*{-2.5in}
\caption[]{The allowed regions in the $\sin \delta$ v $\sin^2 2\theta_{13}$ plane for T2K
experiment,
assuming the true solution is the normal hierarchy with $\sin^2 2\theta_{13} =0.05$ and $\delta = \pi/4$ 
(``{\large $+$}''). 
The upper blue
(lower red) contours are for the normal (inverted) hierarchy
whereas the solid (dashed) contours are for $\cos \delta >0$ $(<0)$. The exposure
is 5 years of both neutrino and anti-neutrino running using a 0.75 MW beam at 3$^o$ off-axis
and HyperKamiokande (30$\times$22.5ktons fiducial mass) as the far detector.
The ellipses correspond to 68, 90 and 99\% C.L. contours.
If the beam intensity is upgraded to 4 MW but only SuperKamiokande is used as the detector the size of the ellipses is significantly increased.}
\end{center}
\end{figure}

\clearpage

\begin{figure}[h]
\begin{center}
\includegraphics[width=4in]{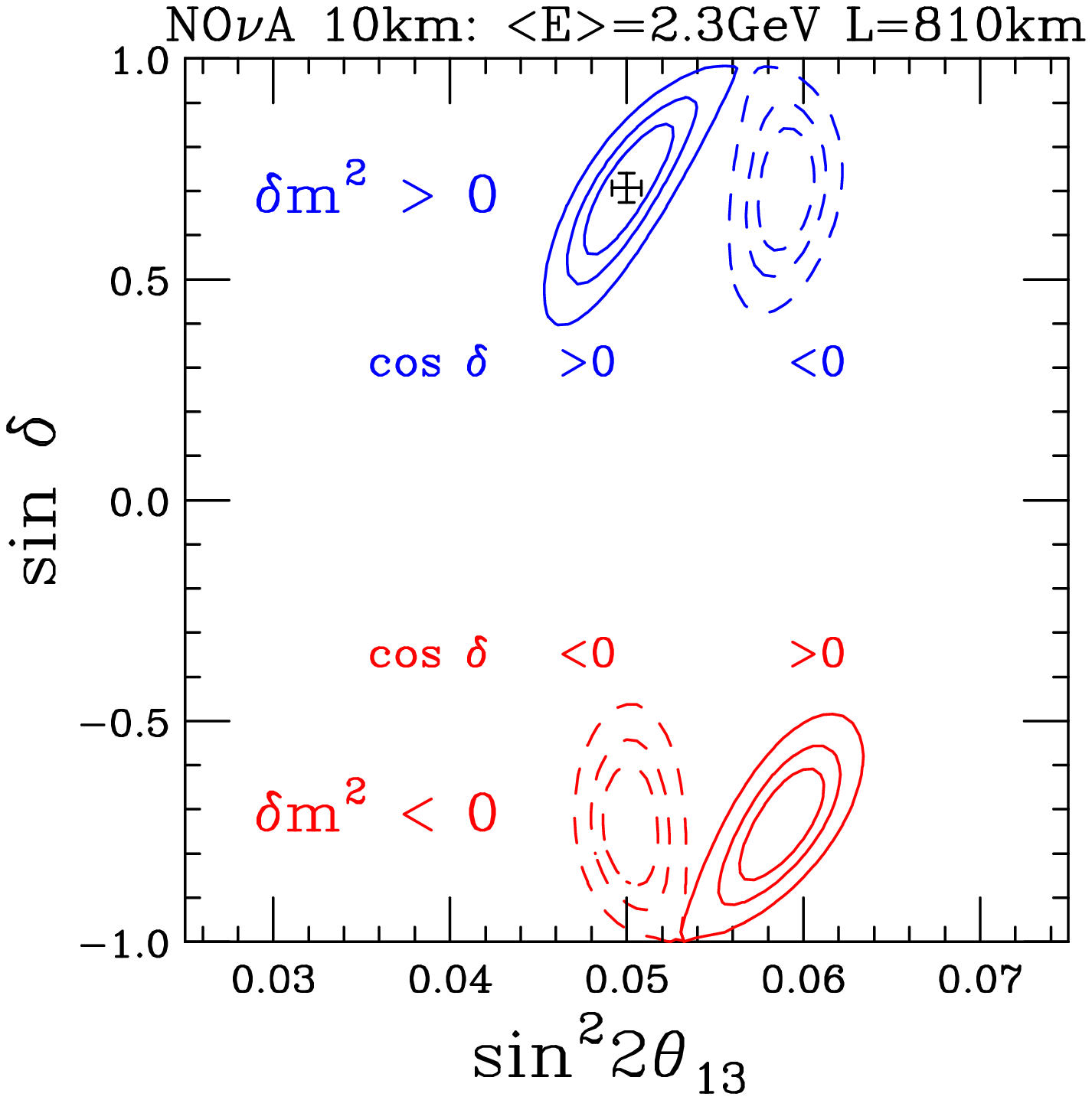}
%\vspace*{-2.5in}
\caption[]{The allowed regions in the $\sin \delta$ v $\sin^2 2\theta_{13}$ plane for the NO$\nu$A
experiment,
assuming the true solution is the normal hierarchy with $\sin^2 2\theta_{13} =0.05$ and $\delta = \pi/4$ (``{\large $+$}''). 
The upper blue
(lower red) contours are for the normal (inverted) hierarchy
whereas the solid (dashed) contours are for $\cos \delta >0$ $(<0)$. The exposure
is 5 years of both neutrino and anti-neutrino running using a 2 MW beam at 10 km off-axis
and 50 kton low Z detector.
The ellipses correspond to 68, 90 and 99\% C.L. contours.}
\end{center}
\end{figure}

\end{document}